\documentclass[journal=jacsat,manuscript=article]{achemso}

\usepackage{txfonts}
\usepackage[T1]{fontenc}
\usepackage{natbib}

\author{Daler R. Dadadzhanov}
\affiliation{ITMO University, 49 Kronverksky Ave., 197101, St. Petersburg, Russia}
\alsoaffiliation{Electrooptics and Photonics Engineering Department, Ben-Gurion University, Beer-Sheva 8410501, Israel}

\alsoaffiliation{Ilse Katz Institute for Nanoscale Science \& Technology, Ben-Gurion University, Beer-Sheva 8410501, Israel}

\email{daler.dadadzhanov@gmail.com}

\author{Tigran A. Vartanyan}
\affiliation{ITMO University, 49 Kronverksky Ave., 197101, St. Petersburg, Russia}

\author{Peter S. Parfenov}
\affiliation{ITMO University, 49 Kronverksky Ave., 197101, St. Petersburg, Russia}

\author{Mikhail A. Baranov}
\affiliation{ITMO University, 49 Kronverksky Ave., 197101, St. Petersburg, Russia}

\author{Alina Karabchevsky}
\affiliation{Electrooptics and Photonics Engineering Department, Ben-Gurion University, Beer-Sheva 8410501, Israel}
\alsoaffiliation{Ilse Katz Institute for Nanoscale Science \& Technology, Ben-Gurion University, Beer-Sheva 8410501, Israel}

\email{alinak@bgu.ac.il}

\title {Broadband plasmonic nanoparticles: fabrication, optical properties, and implications in liquid light chemiluminescence enhancement}


\begin{document}

%
%
%

\begin{abstract}
 Chemiphores are entities, which exhibit wide-band light emission without any external light source but just due to the chemical reaction resulting in the chemiluminescence effect. Since the chemiphores usually have low quantum efficiency, chemiluminescence is a weak optical effect. We found that plasmonic nanoparticles can efficiently enhance the peculiar effect of chemiluminescence due to the acceleration of the radiative decay of the chemiphore excited state which, in turn, enlarges the chemiluminescence yield. Correspondingly, plasmonic nanoparticles are nanoparticles with sub-wavelength sizes experiencing the absorption band in specific wavelength which are characterized by unique optical properties, as well as high localization of electromagnetic radiation. However, the broadband properties of plasmonic nanoparticles and their implications in liquid light, the chemiluminescence effect, is overlooked. Therefore, they can attract attention as novel materials for photonics, sensing, and forensic science. Here, fabrication techniques of broadband plasmonic nanoparticles are reported, and their interesting optical properties together with their applications in chemiluminescence effect are discussed, as well. We fabricated the nanoparticles with laser ablation in liquids (LAL) technique and propose the physical vapor deposition (PVD) synthesis with annealing-assisted treatment for further studies. Both techniques are accessible and allow production of ensembles of nanoparticles having shape and size distributions to exhibit broad plasmonic resonance which fit the wide-band emission of a chemiphore. Our results, in particular, a specific design for plasmonic nanoparticles placed on the dielectric material, lead the way toward a new generation of chemiluminescence-based devices starting from sensing, healthcare, biomedical research and quantum systems such as pump-free laser sources.
\end{abstract}

Chemiluminescence, or liquid light effect, is an emission of photons resulting from an exothermic reaction \cite{Karabchevsky2016TuningNanoparticles}. This fascinating optical effect that finds its use in various applications: from forensic science targeted to detection the blood traces at crime scene \citet{Barni2007ForensicDetection}, to industrial bio-chemistry \cite{Yamashoji2009DeterminationMicroperoxidase,Iranifam2016ChemiluminescenceChemistry}. Due to the emerging need of securing societies, ensuring the safety of the citizens and protection of freedom a development of the technology to strengthen forensic science has been recently boosted. Luminol is a chemiphore (chemical that exhibits chemiluminescence) which under certain conditions emits blue-glow light \cite{Freeman2011ChemiluminescentDots}. Forensic investigators use Luminol to detect trace amounts of blood at crime scenes since it reacts with the iron in hemoglobin; biologists use it in cellular assays to detect copper, iron and specific proteins \cite{Klopf1983EffectPeroxide}. However, the quantum efficiency of Luminol is very low, which significantly limits its potential applications.\

One of the experimental tools for observing the chemiluminescence effect is a microfluidic chip, which is employed for the efficient study of low volume samples by isolating key phenomena from influential surrounding environment \cite{Wei2006ChemiluminescenceFlour}. The mixing during the flow-injection in microfluidic chips helps in obtaining the efficient chemical reaction in which the chemiphores and their oxidants are mixed to emit light. Crucial developments for the fabrication approaches for the microfluidic technology platform already gave birth to emerging applications such as the lung-on-a-chip platform \cite{Perestrelo2015MicrofluidicMicroengineering,Valencia2012MicrofluidicNanoparticles,Aziz2017TheSimulations}, single-cell functional proteomics \cite{Lazar2006MicrofluidicScreening,Yu2014Microfluidics-basedApplications}, and measuring the dynamics of green fluorescence protein \cite{Valero2008GeneDevice,Bennett2009MicrofluidicCells}. However, the efficient enhancement of poor quantum emitters, such as the chemiphore Luminol, has never been proposed. This efficient emission of Luminol is expected to find a purpose in forensic science-on-a-chip, such as an accurate DNA profiling with active lasing dust.\

In this paper we report on the repertoire of efficient Luminol enhancement possibilities within frontiers of nanotechnology and quantum mechanics while enhancing the interaction of plasmon-Luminol due to the fabrication-tuned the wide-band plasmonic resonant absorption of the nanoantennas. It is important to note that he quantum efficiency of the chemiluminescence process is quite low due to the competition with a multitude of non-radiative decay processes \cite{Lee1972QuantumSolvents}. Indeed, radiative decay acceleration of the chemiphore excited state is a reliable way to enhance the chemiluminescence yield that attracts attention of many scientists \cite{Karabchevsky2016TuningNanoparticles,Fobel2014PaperDigital,Ou2007CatalyticChemiluminescence}. The surface plasmon resonance in metallic nanoparticles \cite{Karabchevsky2009TheoreticalEnvironment,Karabchevsky2011Dual-surfaceNanoslits} can be employed for the acceleration of these radiative transitions. To maximize of the chemiphores-nanoparticles interaction, metal nanoparticles have to be placed at an optimum distance from the chemiphores \cite{Khurgin2007EnhancementMerit}. The shape and material of the particles have to be accurately chosen to ensure the overlapping of their plasmonic band with the emission band of the chemiphores. Several studies have been performed to investigate the use of plasmons for enhancing the radiative decay in a microfluidic chip \cite{Aslan2009Metal-enhancedCentury}; however, those researchers used non-resonant nanoparticles. New proof of concept experiment was carried out by using a microfluidic chip \cite{Karabchevsky2016TuningNanoparticles}, wherein we utilized the commercially available nanoparticles with non-optimized spectral location of the plasmon absorption band. We attribute the possibility of enhanced chemiluminescence to the Purcell effect. However, the chemiluminescence is limited in efficiency because excited molecules can lose their energy through non-radiative processes, such as internal conversion and intersystem crossing. An excited molecule can decay either radiatively at the rate $\gamma_R$ or non-radiatively at the rate $\gamma_{NR}$. Therefore, the chemiluminescence enhancement with these nanoparticles is not sufficient and alternative fabrication techniques are needed to engineer the wide-band and tunable plasmonic resonances for the efficient interaction between the nanoparticle acceptors and the chemiphore donors.\

Overall, for effective interaction, between the plasmonic nanoparticle and chemiphore, two conditions should be met: 1) the plasmon band has to overlap in spectrum with the chemiluminescence emission band \cite{Park2017EmergingPhotoluminescence} and 2) chemiphore molecules are to be brought in the close proximity of metal nanoparticles \cite{Geddes2017SurfaceFluorescence}. In this work, we explore the possibility of fulfilling both conditions with silver nanoparticles as \textit{acceptors} and the chemiphore molecules as \textit{donors} of energy quanta. As the plasmon band of a silver nanosphere lies on the far blue wing of the Luminol chemiluminescence band, we employed two different techniques to shift the plasmon band in the desired red direction: enlarging of the nanosphere radius and changing of the nanoparticle shape. To keep the transfer of the technology feasible, we rely on the well - established and rather inexpensive techniques of silver nanoparticle production, namely, pulsed laser ablation in liquids  \cite{Simakin2004NanoparticlesEnvironment,Itina2010OnLiquids,Tilaki2007SizeLiquids} and physical vapor deposition on a dielectric substrate. Our proposals are supported by numerical simulations that showed that the plasmon band of silver nanoparticle may be shifted up to optimally overlap with the Luminol chemiluminescence band.\

Our results can provide a gain media, utilizing plasmonic enhancement, as a toolkit for Forensic laboratories. Gain medium or active medium, is capable of amplification of light by utilizing the process of stimulated emission. Carefully designed nanostructures are equipped of both enhancing fundamental interactions and directing signals to an observer or camera. This engineered chemiluminescence-based devices with a gain media will improve detection limits of chemiluminescence for forensic science, facilitate the imaging, and confer new insights in DNA profiling through important physical effects such as amplification process involved in stimulated emission.\

The utilization of chemicals such as 'light source' is one of the major challenges facing photonics today. The revolutionizing nature of our study is in the efficient interaction between Luminol and nanostructures. Outcomes of our research make it possible to take a step forward and increase the accuracy of DNA profiling, increase the efficiency of chemically induced photon emission by developing new engineered nanostructures capable of efficient chemiluminescence and construct a general platform for lasing dust.

\section{Materials and methods}
\subsection{Synthesis of colloidal silver nanoparticles}
Silver nanoparticles have been synthesized using the laser ablation approach. The experimental setup is shown in Fig.\ref{fgr:1}a. The silver plate (99,99$\%$ purity) has been used as a target for the synthesis. Laser ablation was caused by pulsed Nd:YAG solid-state laser (SOLAR Laser Systems). The laser source enables the second harmonic with 532 nm and turned in the Q-switching regime during ablation process. The laser beam, with a diameter of 5 mm, was focused using a lens with focal length of 60 mm through a transparent liquid (deionized water) on the surface of the metal plate. The surface of the liquid remained free, and the thickness of its layer above the surface of the plate was several centimeters. To get broad size distribution of silver nanoparticles the target was irradiated for 15 min. The pulse energy was set to 100 mJ. The pulse duration and the repetition rate were, respectively, 10 ns and 5 Hz, respectively.

\subsection{Characterization}
The morphology studies of the sample with silver nanoparticles synthesized using pulsed laser ablation were performed using an atomic force scanning microscope on Solver PRO-M (NT-MDT) using silicon probes NSG01. Microscopic examination was performed in the tapping (semi-contact) mode. The resonance frequency and curvature radius of the probe in the tapping mode were 283 kHz and 10 nm, respectively. Scanning electron microscope images of supported silver nanoparticles were obtained by means of MERLIN (Carl Zeiss) microscope at 15 kV.\
Extinction spectra of the colloidal solution in a cuvette with the thickness of 1 cm were collected using an SF-56 spectrophotometer (LOMO) with 1 nm resolution in the range of 250 - 800 nm at room temperature. The chemiluminescent spectra of Luminol were recorded by a charge-coupled device (CCD), Lumenera Infinity 2-3C (Lumenera Corporation, 7 Capella Crt. Ottawa, Ontario, Canada).

\subsection{Numerical simulation}
The Finite-Element-Method (FEM) implemented in COMSOL Multiphysics 5.4 software was used to solve Maxwell's equations for stationary problems. We calculated absorption, scattering and extinction cross-sections spectra of silver nanoparticles of different shapes and on different substrates. We included the perfectly matched layer (PML) in the form of a physical sphere domain to absorb all scattered light. We define the thickness of PML as the half of the incident wavelength. The incident field oscillated in the x-direction (parallel to the substrate) with a k-vector in the -z direction (perpendicular to the substrate). The absorption cross section ($\sigma_{abs}$) was calculated by integrating the power loss density over the particle volume. The scattering cross-section ($\sigma_{sc}$) was derived by integrating the Poynting vector over an imaginary sphere around the particle. The total extinction cross-section was calculated as the sum of the both $\sigma_{ext}$ = $\sigma_{abs}$ + $\sigma_{sc}$. The complex refractive index of the silver nanoparticles was obtained from the experimental work of Johnson and Christy \cite{Johnson1963OpticalMetals}

\section{Results and Discussion}
\subsection{Experimental and numerical studies for nanoparticle-Luminol system}

\begin{figure}
 \centering
 \includegraphics[height=16 cm ]{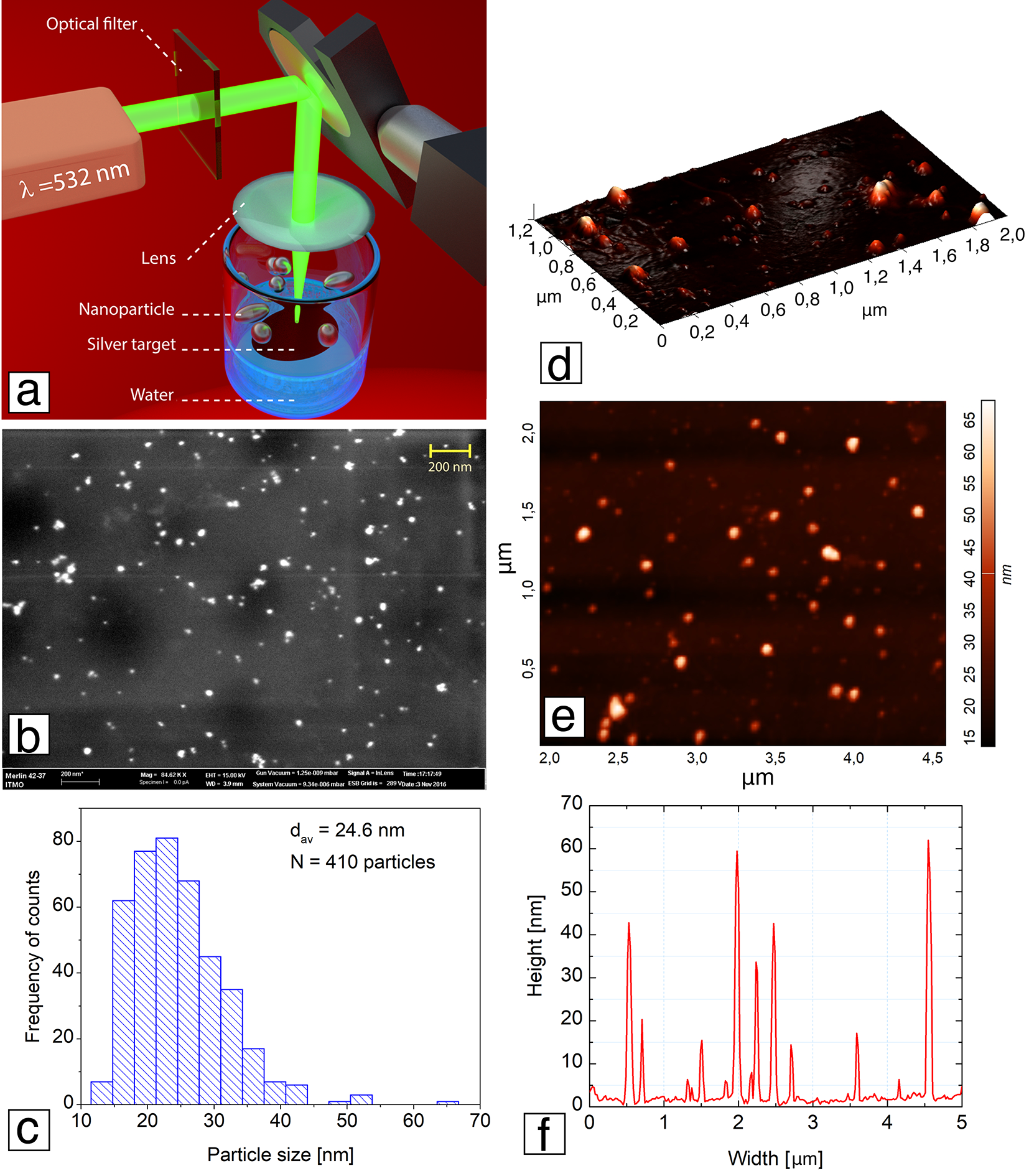}
 \caption{(a) Experimental set up of pulsed laser ablation of silver target in water. The dimensions are not to scale.(b) Scanning electron microscope image of silver nanoparticles on the carbon grid. The scale bar is 200 nm. (c) The size distribution of silver nanoparticles \textit{d$_{av}$} = 24.6 nm.(d) 3D atomic force microscopy image of silver nanoparticles dried on the quartz substrate obtained in the tapping mode regime. (e) atomic force microscopy image (the top view) with size scale bar from 15 to 65 nm. (f) The cross-section profile of fabricated silver nanoparticles. The height of nanoparticles varied from 10 to 60 nm.}
 \label{fgr:1}
\end{figure}

We propose the plasmonic system design for tuning the broad absorption of plasmonic nanoparticles for efficient enhancement of the chemiluminescence effect. To prove the efficiency of the proposed \textit{nanoparticle} + \textit{chemiphore} system, we first constructed the experimental setup shown in Fig.\ref{fgr:1}a and fabricated silver nanoparticles in liquid. During the experiment, performed at the room temperature, the silver target was ablated by the laser for different exposure times. This is done to find the time dependence on the particle size which, in turn, gives the broad enough plasmonic band to enhance the chemiluminescence effect and using the distilled water without a stabilizing agents or surfactants. After fabrication, we characterized the nanoparticles with the atomic force microscope (AFM) and the scanning electron microscope (SEM). Based on the SEM (Fig.\ref{fgr:1}b) micrographs, we estimated the size distribution of nanoparticles. The histogram in Fig.\ref{fgr:1}c confirms that the colloidal solution contains fabricated nanoparticles with a broad size distribution. The average diameter size of silver nanoparticles is 24.6 nm. The measured extinction spectra of colloidal silver nanoparticles exhibit the pronounced plasmon band in spectral range of 300 to 600 nm. Dried silver nanoparticles were imaged by AFM on a quartz substrate (Fig.\ref{fgr:1}d,e). According to AFM data of the cross-section (Fig.\ref{fgr:1}f),  the particle height for a specific region in AFM image varied from h=10 to h=60 nm (taking into account the shape of the AFM probe), while their dimensions in substrate plane were from 10 to 50 nm in diameter.

\begin{figure}
 \centering
 \includegraphics[height=12 cm]{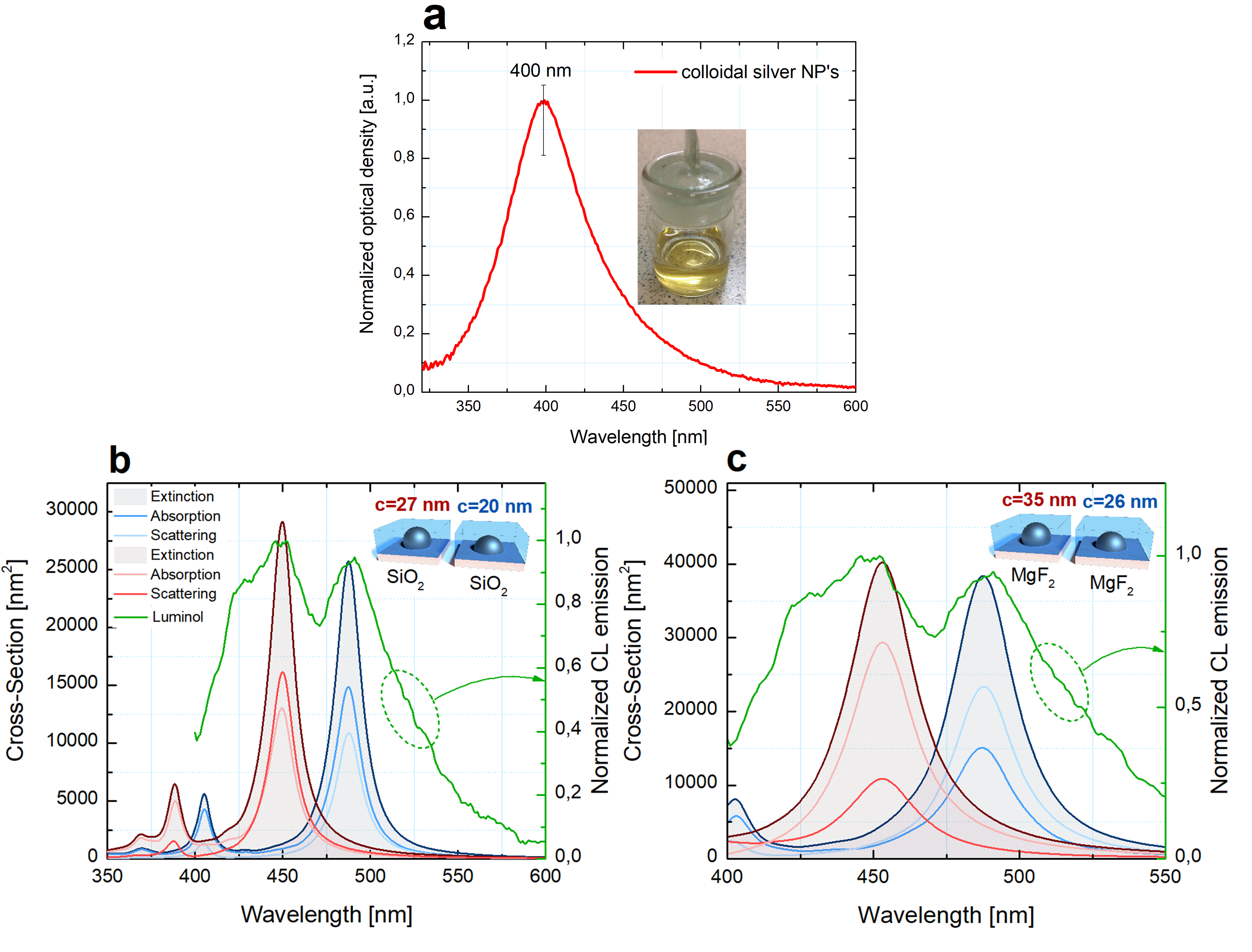}
 \caption{(a) Measured in visible spectrum optical densities of colloidal silver nanoparticles prepared by laser ablation technique in water and normalized to the maximum value. In the inset is the colloidal solution with distinctive yellow color. (b) Overlap between the measured chemiluminescence intensity of the Luminol solution (green line) and the calculated extinction spectra of supported silver nanoparticle submerged into Luminol solution; extinction cross-section spectra of silver nanoparticle with the height of 27 nm for the enhancement of the Luminol chemiluminescence band at 452 nm and with the height of 20 nm for the Luminol band at 489 nm. The diameters of silver nanoparticles in both cases are 40 nm. (c) Extinction cross-section spectra of silver nanoparticles with the heights of 35 nm and 26 nm on the magnesium fluoride substrate. The diameters in both cases are equal to 52 nm. The 3D models of silver nanoparticles placed on the respective substrates and submerged into Luminol solution are illustrated in the inserts.}
 \label{fgr:2}
\end{figure}

Fig.\ref{fgr:2}a shows a photograph of the beaker with fabricated silver nanoparticle in distilled water. We notice that the liquid has definite yellow color. This color can be explained by the spectral location of absorption band of fabricated nanoparticles in the absorption spectra, as shown in Fig\ref{fgr:2}a. Thus, the laser ablation of the silver target lasted 15 minutes leads to the formation of the silver nanoparticles with the broad plasmon band centered at 400 nm.\

For this reason, we addressed the deposition of silver nanoparticle on the dielectric substrate which is more favorable for further plasmon shifting. A much larger long-wavelength shift of the silver nanoparticles plasmon bands can be obtained by physical vapor deposition (PVD) of silver on a transparent dielectric substrate, for example, sapphire, magnesium fluoride (MgF$_2$), quartz (SiO$_2$) or titanium dioxide (TiO$_2$). Fig.\ref{fgr:2}b,c show the numerical calculations of extinction, absorption and scattering cross-sections spectra for nanoparticles on different substrates. During the modeling, the silver nanoparticle was embedded in a Luminol solution and placed on the two types of dielectric material. We show that by changing the size parameter we can tune the optical properties of the silver nanoparticle. Firstly, to find the optimized parameter we measured the chemiluminescent emission spectra of Luminol in water oxidized by hydrogen peroxide. Since the Luminol was diluted with high ratio in water, for calculation, we took for calculation the refractive index of the surrounded media of 1.33. According to our calculations presented in Fig.\ref{fgr:2}b, ideal matching is achieved for silver nanoparticles with the height (denoted as c in Fig.\ref{fgr:2}) of 20 nm and 29 nm deposited on a quartz substrate (the refractive index is 1.46 \cite{Noguez2007SurfaceEnvironment}). In this case the diameter was constant (40 nm). However, the calculations show that the plasmon band of hemispherical silver nanoparticles on a quartz substrate cannot be brought in an exact resonance with the second Luminol chemiluminescence emission peak located at 452 nm. Indeed, even in the hemispheres with the diameter as small as 5 nm plasmonic peak is located at 472 nm. To overcome this difficulty, the substrate of smaller refractive index may be used. Of particular interest is using of a wide-spread material MgF$_2$, which has the refractive index as low as 1.38 \cite{Bartkowiak2016PorousApplications}. In Fig.\ref{fgr:2}c, we show the extinction cross-section spectra of silver nanoparticles with diameters of 52 nm and two different heights: h=35 nm and h=26 nm on the magnesium fluoride substrate. Fig. \ref{fgr:2}c shows that the suitable overlapping between the Luminol emission band and the silver nanoparticle plasmon band is achieved when: a) the hemispherical particles height of h=26 nm is placed on top of the magnesium fluoride substrate and b) the height of the nanoparticle is enlarged up to h=35 nm. We notice that the sapphire substrate having the refractive index as high as 1.77 \cite{Knight2009SubstratesNanoparticle,Liu2015TunableLayers} is not suitable for efficient overlapping of Luminol emission, since the long-wavelength shift is too large even for the very small silver hemispheres. \

\begin{figure}
 \centering
 \includegraphics[height=15 cm]{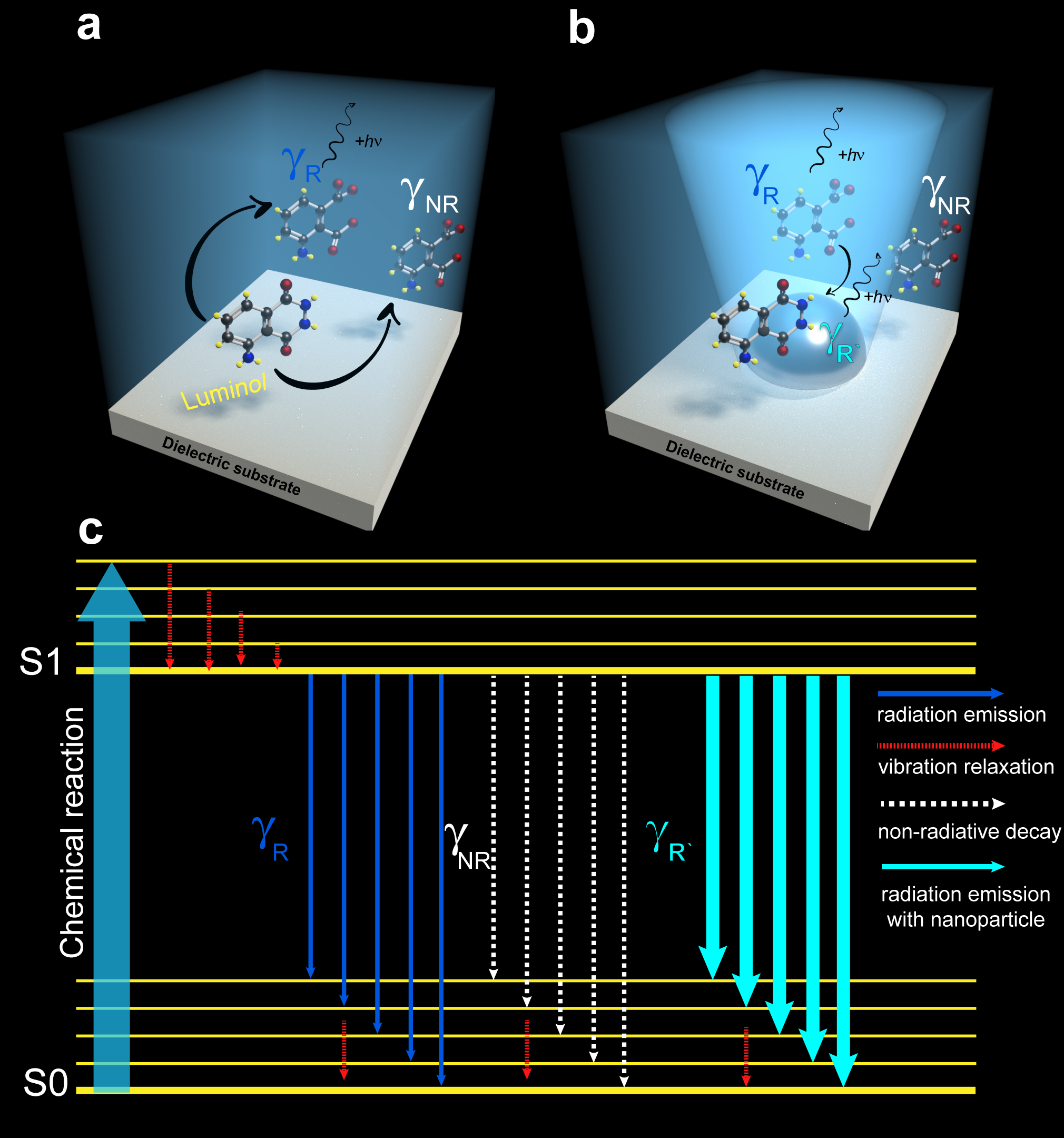}
 \caption{Schematics of the enhancement mechanism of chemiluminescence effect. (a) Oxidized Luminol lead to competitive process of radiative decay (emission of the photon) and non-radiative decay (heat). The blue color emission is attributed to emission of Luminol. (b) The proposed mechanism metal-enhanced chemiluminescence in the presence of silver nanoparticle. (c) The general Jablonski energy level diagram modified for the chemiluminescence emission with the presence of silver nanoparticle \cite{Mirasoli2016ChemiluminescenceBiomedicine}. The chemiluminescence emission origin as result of chemical oxidation process. The molecular transition from the excited (S1) to the electron ground state(S0), emitting a photon and heat.$\gamma_R$ and $\gamma_{NR}$ are the radiative and non-radiative decay of Luminol molecule. $\gamma_{R'}$ is the radiative decay rate with silver nanoparticle.}
 \label{fgr:3}
\end{figure}

Fig. 3 summarizes the schematics of the chemiluminescence enhancement mechanism. Fig.\ref{fgr:3}a represents the Luminol excitation in the course of a chemical reaction with oxidizer followed by two competitive decay routes: radiative and non-radiative. The effect of silver nanoparticles on the Luminol emission is depicted in Fig.\ref{fgr:3}c. The expected chemiluminescence enhancement can be explained by coupled emission from excited states of Luminol and plasmon mode in silver nanoparticles \cite{Aslan2009Metal-enhancedCentury}.

Indeed, as it was reported in previous works \cite{Hicks2005EM2005.,Aslan2009SurfaceChemiluminescence}, in contrast of the metal-enhanced fluorescence (MEF) \cite{Karabchevsky2012MicrospotFilms,Abdulhalim2009Surface-enhancedWater} where the molecule initially must be pumped by external light, metal-enhancement chemiluminescence (MEC) origins stem by a chemical oxidation process as shown in the Jablonski diagram (Fig.\ref{fgr:3}c). Therefore, the decay rate can be increased by plasmon nanoparticle.\

We experimentally prove the concept of operation and feasibility of the engineered optimized system. First, we fabricated the silver nanoparticles on the quartz substrate using the physical vapor deposition. Silver target was sputtered in high vacuum at the rate of 0.1 \AA/s and then annealed at 200$^o$ C. Second, we controlled the spectral position of the plasmon absorption band by the amount of the sputtered metal measured as an equivalent thickness of the continuous film. We found that the maximum of the plasmon absorption band is located at 450 nm when the equivalent thickness is 3 nm. However, the plasmon band shifts to 488 nm when the equivalent thickness is 9 nm \cite{Toropov2017FabricationOscillations,Leonov2015EvolutionAnnealing} according to the requirements of the effective acceleration of radiative transitions in Luminol. We conclude that the fabrication of silver nanoparticles using the physical vapor deposition technique in vacuum followed by the self-organization of nanoparticles is an affordable and convenient method that potentially can lead to plasmonic enhancement for the Luminol chemiluminescence.\

In the above discussion we were mainly concerned with the spectral overlap of the plasmon and chemiluminescence bands. Now we address the second requirement of placing the chemiphore molecules in the close proximity of metal nanoparticles. One can underline that the arrangement of particles on the transparent dielectric substrate makes it possible to control the distance between plasmonic nanoparticles and chemiphores using, for example, a polymer thin layer as a spacer. Another advantage is the possibility of incorporating the substrates modified with plasmonic nanoparticles in a microfluidic chip for chemiluminescence enhancement. The width as well as the position of plasmon bands of nanoparticles obtained via PVD and subsequent annealing can be readily tuned to match the position and the width of the chemiphore bands.
The outcomes of our research leading to the chemiluminescence enhancement are very important in chemi- and bio- luminescence effects for variety of applications in biology, sensing, healthcare and forensic science. 
\subsection{Numerical calculation for alternative MEC systems on dielectric substrate}

\begin{figure}
 \centering
 \includegraphics[height=5 cm]{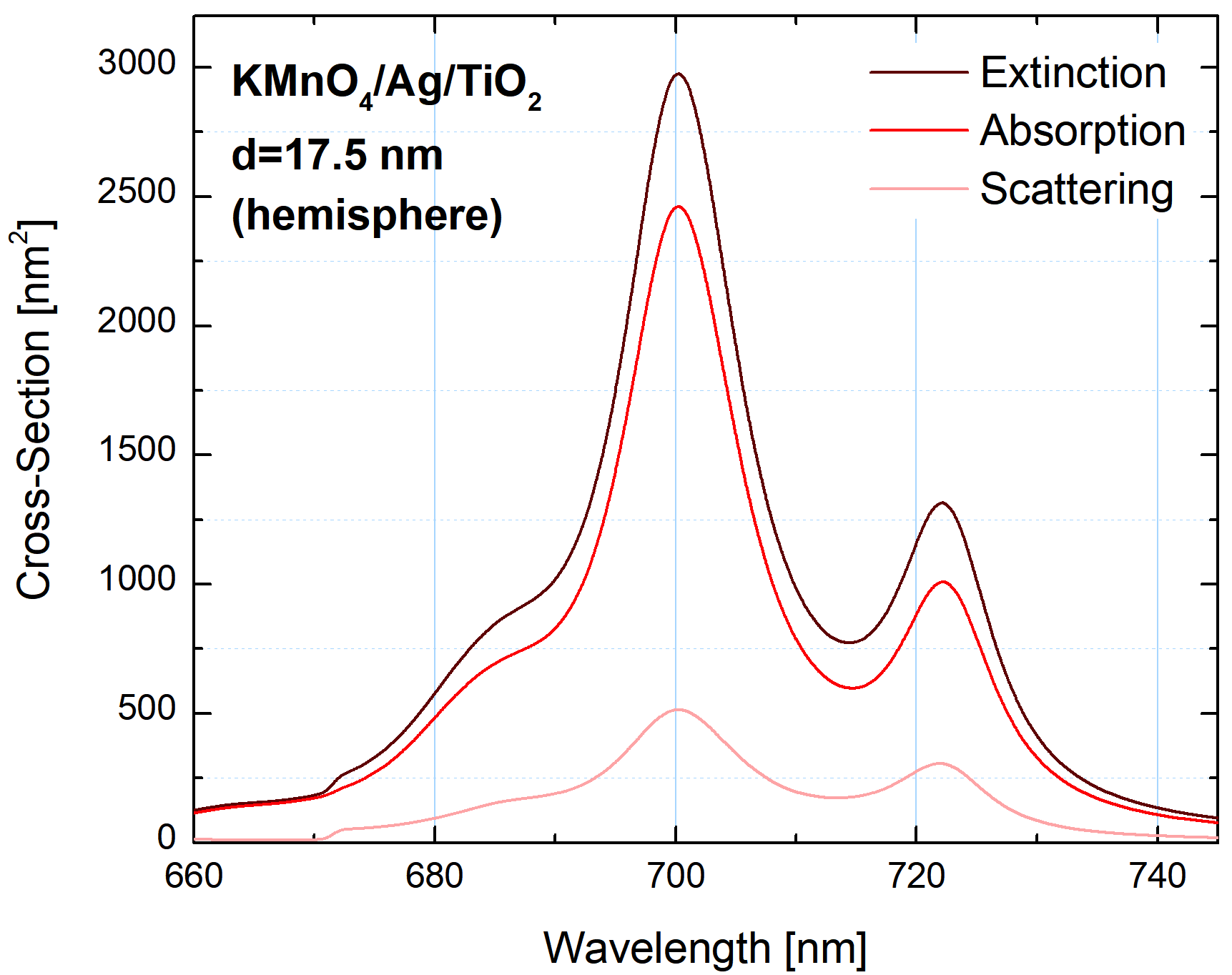}
 \caption{Calculated absorption, scattering and extinction cross-section spectra of the silver nanoparticle. The silver nanoparticle deposited on the TiO$_2$ substrate. The diameter of hemisphere is d=17.5 nm.}
 \label{fgr:4}
\end{figure}

\begin{figure}
 \centering
 \includegraphics[height=16 cm]{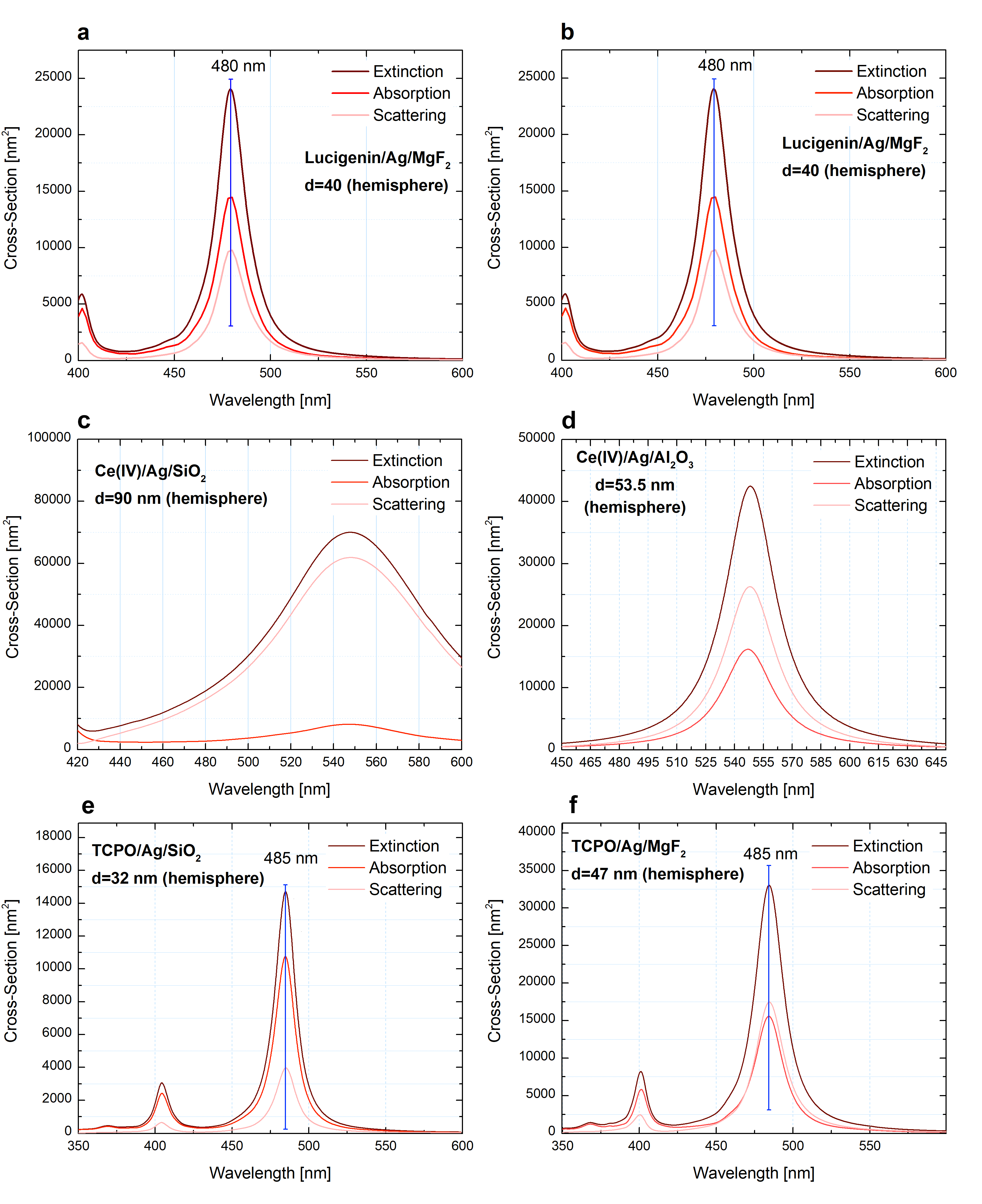}
 \caption{Calculated absorption, scattering and extinction cross-section spectra of the silver nanoparticle: (a)deposited on the SiO$_2$ substrate for Lucigenin. The diameter of hemisphere is d=28 nm. (b) the silver nanoparticle on the MgF$_2$ substrate with diameter of 49 nm for Lucigenin system; (c) deposited on the SiO$_2$ substrate for Ce (IV). The diameter of hemisphere is d=90 nm. (d) the silver nanoparticle on the Al$_2$O$_3$ substrate (d=53.5 nm);  (e)deposited on the SiO$_2$ substrate for TCPO system. The diameter of hemisphere is 32 nm. (f) the silver nanoparticle on the MgF$_2$ substrate with diameter of 47 nm for overlapping with TCPO chemiphore.}
 \label{fgr:5}
\end{figure}

As mentioned before, in the present work, we consider the chemiphore Luminol, however here we show designs for the enhanced chemiluminescence of several well-known chemiphores. On the other hand, emission of different chemiphores may be enhanced via the suggested technique and the dedicated choices of substrates and annealing procedures as well \cite{Toropov2017FabricationOscillations}. We found a number of substrate-nanoparticle pairs optimized for surface enhanced chemiluminescence of several well-known chemiphore: KMnO$_4$ in Fig.\ref{fgr:4}, Lucigenin in Fig.\ref{fgr:5} a,b, Ce (IV) in Fig.\ref{fgr:5} c,d,  and bis-(2,4,6-tricholorophenyl) oxalate (TCPO) in Fig.\ref{fgr:5} e,f.\ 

We optimized the cross-section spectra of the silver hemisphere for KMnO$_4$ chemiluminescent emission spectra (Fig.\ref{fgr:4}). The maximum emission peak is 700 nm \cite{Hindson2010MechanismChemiluminescence}. An alternative example of a chemiluminescent agent we considered Lucigenin. This chemiphor is widely used in analytical science and has a maximum emission peak of 480 nm. \cite{He2012FabricationChemiluminescence,Guo2007LucigeninAdsorbates,Okajima2003ChemiluminescenceSolutions,Liu2011PlatinumChemiluminescence} (Fig.\ref{fgr:5}a,b). The cross-section spectra of silver hemisphere optimized for Cerium (IV) chemiluminescent emission spectra presented in Fig.\ref{fgr:5}c,d. The maximum emission peak ~ 550 nm \cite{Yu2012ANanoclusters,Li2009EnhancedNanoparticles}. Next, we calculated the suitable parameters for the plasmon nanoparticle on the top of dielectric substrate for TCPO chemiluminescence emission spectra. The maximum emission peak is 485(Fig.\ref{fgr:5}e,f) nm \cite{Wang2011DeterminationNanoparticles}.

\section{Conclusions}

In summary, we proposed revolutionizing new approaches in designing new materials assembled from nanoparticles and chemiphores for enhancing effects such as of chemiluminescence as well as spasing for emerging applications such as DNA profiling and forensic studies - all those, hold a promise to deliver substantial enhancement in specific and accurate detection, which is not possible via traditional approaches. We noticed that the red edge of the plasmon band located at 400 nm overlaps with the Luminol emission bands. We found that the arrangement of particles on the surface of transparent dielectrics makes it possible to control the distance between the light-emitting species and nanoantennas to increase the Purcell effect of the materials include arrays of sub-wavelength particles, which vary in their shape and size. Combining these materials with recent advances in chemiluminescence enhancement effect demonstrated using flow-injection systems opens the door for new chemiphore-based light sources. These light sources in which the poor quantum emitter, chemiphore, is replaced by high quantum efficiency of plasmonic nanoantenna could not have been achieved by conventional manufacturing processes in which particles size and shape are fixed.  

\section*{Acknowledgement}
The work was supported by the State of Israel- Innovation Authority, Ministry of Economy Grant No. 62045; the Ministry of Science and Higher Education of Russian Federation (Project 3.4903.2017/6.7). This work also was financially supported by Government of Russian Federation, Grant 08-08. The research described was performed as part of a joint Ph.D. program between the BGU and ITMO.

\bibliography{ArXiv}

\end{document}